\documentclass[aps,prc,twocolumn,superscriptaddress]{revtex4-1}
\usepackage{graphicx}
\usepackage{amsmath}
\usepackage{subfigure}
\usepackage{subfigure,dcolumn}
\usepackage{appendix}
\usepackage{xcolor}
\usepackage{ulem}
\usepackage[colorlinks,
linkcolor=blue,
anchorcolor=blue,
urlcolor=red,
citecolor=blue]{hyperref}
\begin{document}
	
	
\title{Dynamical Coulomb effects on charged-pion HBT-radius splitting in central Au+Au collisions at few-GeV energies}
	

\author{Pengcheng Li}
\affiliation{School of Science, Huzhou Normal University, 313000 Huzhou, China}
\affiliation{Bogoliubov Laboratory of Theoretical Physics, Joint Institute for Nuclear Research, 141980 Dubna, Russia}
\author{Yongjia Wang}
\affiliation{School of Science, Huzhou Normal University, 313000 Huzhou, China}
\author{Qingfeng Li}
\email[Corresponding author, ]{liqf@huznu.edu.cn}
\affiliation{School of Science, Huzhou Normal University, 313000 Huzhou, China}

\date{\today}

\begin{abstract}		

The pair transverse momentum ($k_{T}$) and collision-energy ($\sqrt{s_{NN}}$) dependent difference between $\pi^{+}\pi^{+}$ and $\pi^{-}\pi^{-}$ femtoscopic correlations was observed by the HADES and STAR Collaborations. We investigate how much of the observed splitting can be accounted for by Coulomb dynamics in 0–10\% central Au+Au collisions at $\sqrt{s_{NN}}=2.42$–5.2 GeV. Using the UrQMD model and the CRAB programme, three transport scenarios that successively include baryon–baryon, baryon–meson, and meson–meson Coulomb interactions are considered. In a separate calculation based on the baryon–baryon reference, we reconstruct the time-dependent effective charge $Z_{\mathrm{eff}}(t)$ and radius $R_{\mathrm{eff}}(t)$ of the residual charged source and propagate each pion from its individual last strong-interaction point through the evolving field. The calculations reproduce the main $k_{T}$ and $\sqrt{s_{NN}}$ dependences of the radii. Baryon--meson Coulomb interactions generate most of the additional charge splitting in the microscopic calculation. The residual-source treatment also enhances the longitudinal and sideward radius ratios relative to the reference, whereas the outward ratio changes little, and the enhancement generally decreases with increasing $\sqrt{s_{NN}}$. Within the present framework, this contribution does not fully account for the data, motivating further investigation of the charged-source geometry and its interplay with strong-interaction dynamics.

\end{abstract}


\maketitle
	
\section{Introduction}\label{section1}

The Hanbury Brown--Twiss (HBT) method originated as intensity interferometry for measuring stellar angular diameters \cite{Brown:1956zza}. 
Its application to identical pions established that Bose--Einstein correlations at small relative momentum encode the space--time structure of particle emission \cite{Goldhaber:1960sf}. 
In the longitudinally comoving system (LCMS), the extracted HBT radii $R_{\mathrm{out}}$, $R_{\mathrm{side}}$, and $R_{\mathrm{long}}$ constrain complementary aspects of the transverse geometry, emission duration, longitudinal expansion, and source lifetime \cite{Zajc:1984vb,Pratt:1986cc,Bertsch:1988db,Rischke:1996em,Lisa:2005dd}. 
The collision energy and pair-transverse-momentum dependences of these radii and ratios have therefore been used to investigate the high-density equation of state (EoS), reaction dynamics, and possible nonmonotonic behavior associated with the QCD phase diagram \cite{Pratt:1986cc,Rischke:1996em,Li:2008qm,Li:2022iil,Jiang:2026pox}.

Measurements by HADES and STAR at collision energies of a few GeV show that the radii extracted from $\pi^-\pi^-$ pairs exceed those from $\pi^+\pi^+$ pairs, most prominently for the sideward and longitudinal components \cite{HADES:2019lek,Luong:2024eaq,Luong:2026ugh}. 
The splitting is largest at low pair transverse momentum and generally decreases with increasing collision energy. 
A quantitative interpretation remains unsettled because the fitted radii depend on several coupled aspects of the collision dynamics, including the initial nuclear structure, the neutron-rich isospin composition, the EoS and mean-field dynamics, resonance production and absorption, and Coulomb interactions \cite{Li:2007yd,Fang:2022kru,Li:2022icu,Li:2025mox,Kincses:2025iaf,Luong:2026ugh,Khyzhniak:2026skh,Xi:2026vrp,Li:2007im}. 
Interpreting the measured splitting therefore requires consideration of these competing contributions \cite{Khyzhniak:2026skh}.

In some literature, analytical treatments emphasised the Coulomb interaction between an emitted pion and the positively charged matter remaining in the collision zone \cite{Barz:1996gr,Barz:1997es,Shoppa:1998sw,Lednicky:2005tb}. 
A recent STAR analysis showed that a compact description based on an effective charge at each collision energy and a charge-dependent momentum correction can account for the measured HBT-radius splitting \cite{Luong:2026ugh}. 
This successful phenomenological description motivates a complementary transport treatment that resolves the different Coulomb-field histories sampled by pions emitted at different space-time points.

In heavy-ion collisions, pion kinetic freeze-out occurs over an extended time interval \cite{Li:2007im,Li:2022iil}.
An early-emitted pion encounters a compact source, whereas a late-emitted pion probes a more dilute and spatially extended charge distribution. 
Moreover, the source continues to evolve as the pion propagates. 
Consequently, the accumulated Coulomb impulse depends on the pion charge, freeze-out space-time point, momentum direction, and subsequent trajectory \cite{Sinyukov:1998fc,Maj:2009ue,Cebra:2014sxa}. 
A dynamical treatment must specify which part of this interaction history is included in transport and which part is represented by subsequent propagation through an effective field.

In this work, we use UrQMD and CRAB to quantify the extent to which Coulomb dynamics account for the observed charge splitting \cite{Bass:1998ca,Bleicher:1999xi,Pratt:1990zq}. 
Three transport scenarios successively include baryon--baryon, baryon--meson, and meson--meson Coulomb interactions. 
In a complementary calculation, the residual charged source is represented by an expanding uniform sphere characterized by the time-dependent quantities $Z_{\mathrm{eff}}(t)$ and $R_{\mathrm{eff}}(t)$. 
Each charged pion is propagated from its individual last strong-interaction point through the corresponding Coulomb field. 
The aim is to assess whether a dynamical effective-source description can account for most of the observed splitting, as suggested by the phenomenological analysis. 
We quantify the Coulomb contribution to the $\pi^-\pi^-$--$\pi^+\pi^+$ HBT-radius splitting, while systematic variations of nuclear structure, isospin-dependent dynamics, and mean-field effects are left for future work.

\section{Methodology}\label{sec:model}
\subsection{UrQMD model and two-body Coulomb scenarios}

In this work, Au+Au collisions at $\sqrt{s_{NN}}=2.42$, 3.0, 3.2, 3.5, 3.9, 4.5, and 5.2 GeV are simulated using UrQMD version 4.0 \cite{Bass:1998ca,Bleicher:1999xi}. 
We employ a hard Skyrme EoS with an incompressibility of $K_0=380$ MeV and no momentum-dependent potential. 
Hard-EoS transport calculations have also been employed in studies of collective flow and two-pion interferometry in the few-GeV regime~\cite{Hillmann:2018nmd,Lan:2022rrc,Li:2022iil,STAR:2021yiu}. 
The initial isospin composition, EoS stiffness, in-medium cross sections, and other transport-model ingredients are kept fixed throughout this study.

Three two-body Coulomb scenarios are considered. 
Case I includes only baryon--baryon (BB) Coulomb interactions and serves as the reference. 
Case II additionally includes baryon--meson (BM) interactions, and Case III further includes meson--meson (MM) interactions.
The baryonic sector comprises charged nucleons, $\Delta$ and $N^*$ resonances, hyperons and hyperon resonances, and the corresponding antibaryons. 
The mesonic sector comprises charged pseudoscalar and vector mesons and their resonances, including $\pi^{\pm}$, $K^{\pm}$, and $\rho^{\pm}$. 
Cases I--III are compared with the residual-source field disabled. 
Conversely, the residual-source calculation is applied to the Case-I phase-space output, in which pions do not experience direct BM or MM Coulomb forces. 
This design avoids double counting the Coulomb interaction with the surrounding charged medium at the microscopic and effective-source levels.

Case III and the residual-source calculation are complementary rather than equivalent implementations of Coulomb dynamics. 
In Case III, pairwise Coulomb forces are evaluated event by event during transport. They can modify pion trajectories, subsequent scattering histories, and the position--momentum correlations at freeze-out. 
The residual-source calculation starts at each pion's last strong-interaction point. 
It represents the surrounding charges by an event-averaged spherical monopole. 
This effective source is specified by the mean net charge and a radius derived from the charge-weighted root-mean-square (RMS) size \cite{Gyulassy:1980xb,Baym:1996wk,Hardtke:1997cy}. 
Thus, these two calculations are compared as distinct dynamical descriptions and are not combined additively.

\subsection{Reconstruction of $Z_{\mathrm{eff}}(t)$ and $R_{\mathrm{eff}}(t)$}
\label{subsec:source}

\begin{figure}[t!]
\centering
\includegraphics[width=0.45\textwidth]{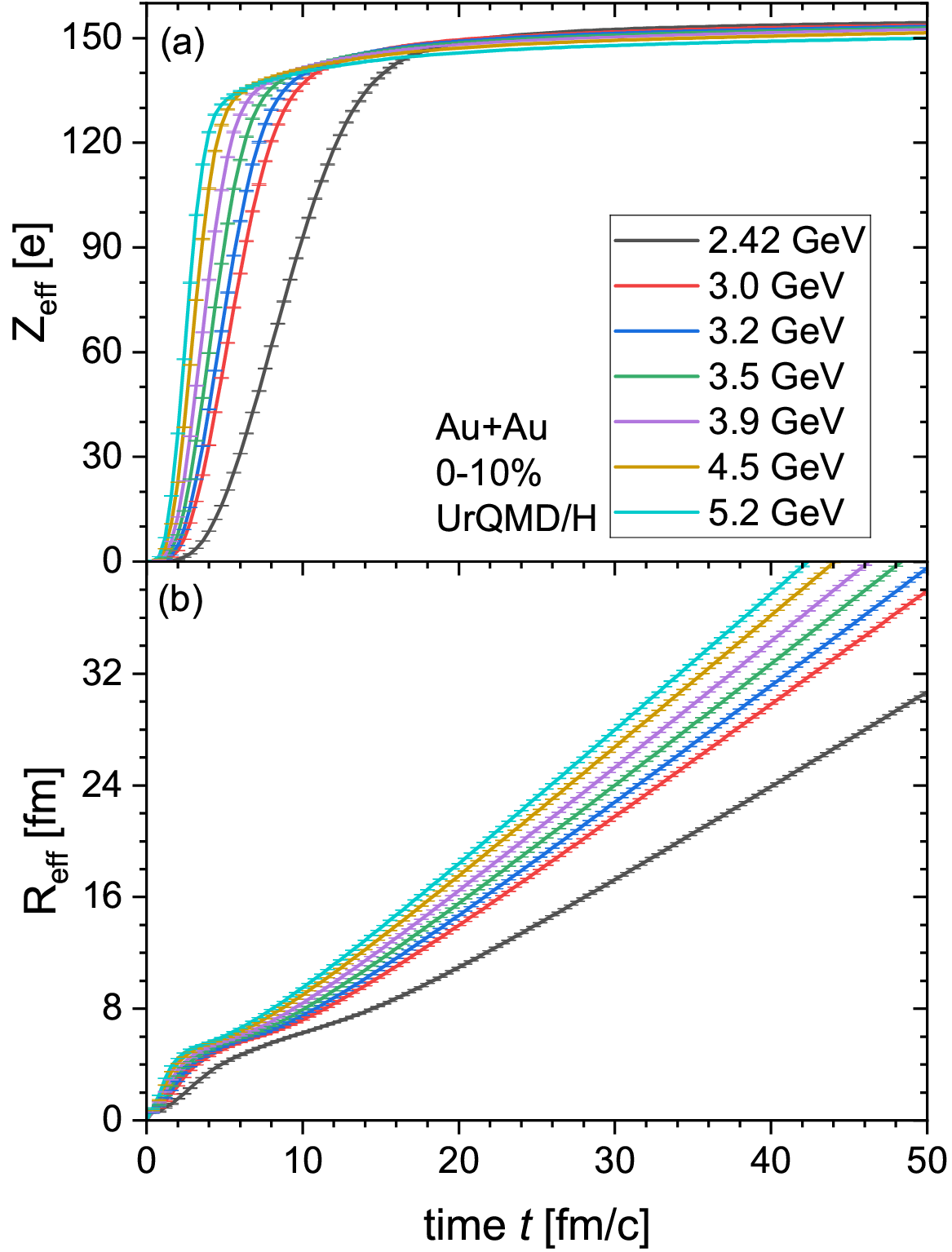}
\caption {\label{fig:zeff-reff-time}(Color online) Time evolution of the effective charge $Z_{\mathrm{eff}}(t)$ (a) and the equivalent uniform-sphere radius $R_{\mathrm{eff}}(t)$ (b), reconstructed from UrQMD simulations of 0--10\% central Au+Au collisions. Error bars represent the standard errors of the event means.}
\end{figure}

The residual charged source is reconstructed from time-dependent UrQMD phase-space snapshots sampled at intervals of $0.4$ fm/$c$ from $t=0$ to 50 fm/$c$.
At each collision energy, $5\times10^5$ events are analyzed. 
An incident baryon is classified as a participant after undergoing at least one elastic or inelastic scattering. 
For event $e$, we define the effective residual-source ensemble $\mathcal{S}_e(t)$ as the charged participant baryons together with all charged mesons present in the snapshot at time $t$. 
Unscattered projectile and target baryons are classified as spectators and are excluded. 
This participant-dominated definition is motivated by the small spectator contribution expected near midrapidity in central collisions \cite{HADES:2022mwn}. 

For event $e$ at time $t$, the net source charge $Z_e(t)$ is $\sum_{a\in\mathcal{S}_e(t)}q_a$, where $a$ labels a particle in the residual source, $q_a=Q_a/e$ is its signed electric charge in units of the elementary charge $e$.
The charge centroid and the signed-charge-weighted three-dimensional RMS radius are defined as
\begin{align}
 \boldsymbol{R}_{\mathrm{ch},e}(t)&=
 \frac{1}{Z_e(t)}\sum_{a\in\mathcal{S}_e(t)}q_a\boldsymbol{r}_a(t), \\
 r_{\mathrm{rms},e}^{2}(t)&=
 \frac{1}{Z_e(t)}\sum_{a\in\mathcal{S}_e(t)}q_a
 \left|\boldsymbol{r}_a(t)-\boldsymbol{R}_{\mathrm{ch},e}(t)\right|^2.
 \label{eq:event-rms}
\end{align}
For each event, we map the charge distribution onto a uniform sphere with the same signed-charge second moment. 
The sphere retains the time-dependent net charge and a characteristic charge radius while providing a finite, continuous Coulomb field inside the source. 
Because $r_{\mathrm{rms}}=\sqrt{3/5}\,R$ for a uniform sphere of sharp radius $R$, the effective radius $R_{\mathrm{eff},e}(t)$ is $\sqrt{5/3}r_{\mathrm{rms},e}(t)$. 

The effective-source parameters are the event means,
\begin{equation}
 Z_{\mathrm{eff}}(t)=\langle Z_e(t)\rangle_e,\qquad
 R_{\mathrm{eff}}(t)=\langle R_{\mathrm{eff},e}(t)\rangle_e .
\end{equation}
The resulting $Z_{\mathrm{eff}}(t)$ and $R_{\mathrm{eff}}(t)$ are shown in Fig.~\ref{fig:zeff-reff-time}. The effective charge rises rapidly during the early stage and subsequently approaches a plateau. 
The more rapid charge build-up and radial expansion at higher collision energies tend to shorten the interval during which the residual source remains compact and generates a strong Coulomb field. 

\subsection{Time-dependent residual-source Coulomb field}
\label{subsec:afterburner}

\begin{figure}[t]
\centering
\includegraphics[width=0.45\textwidth]{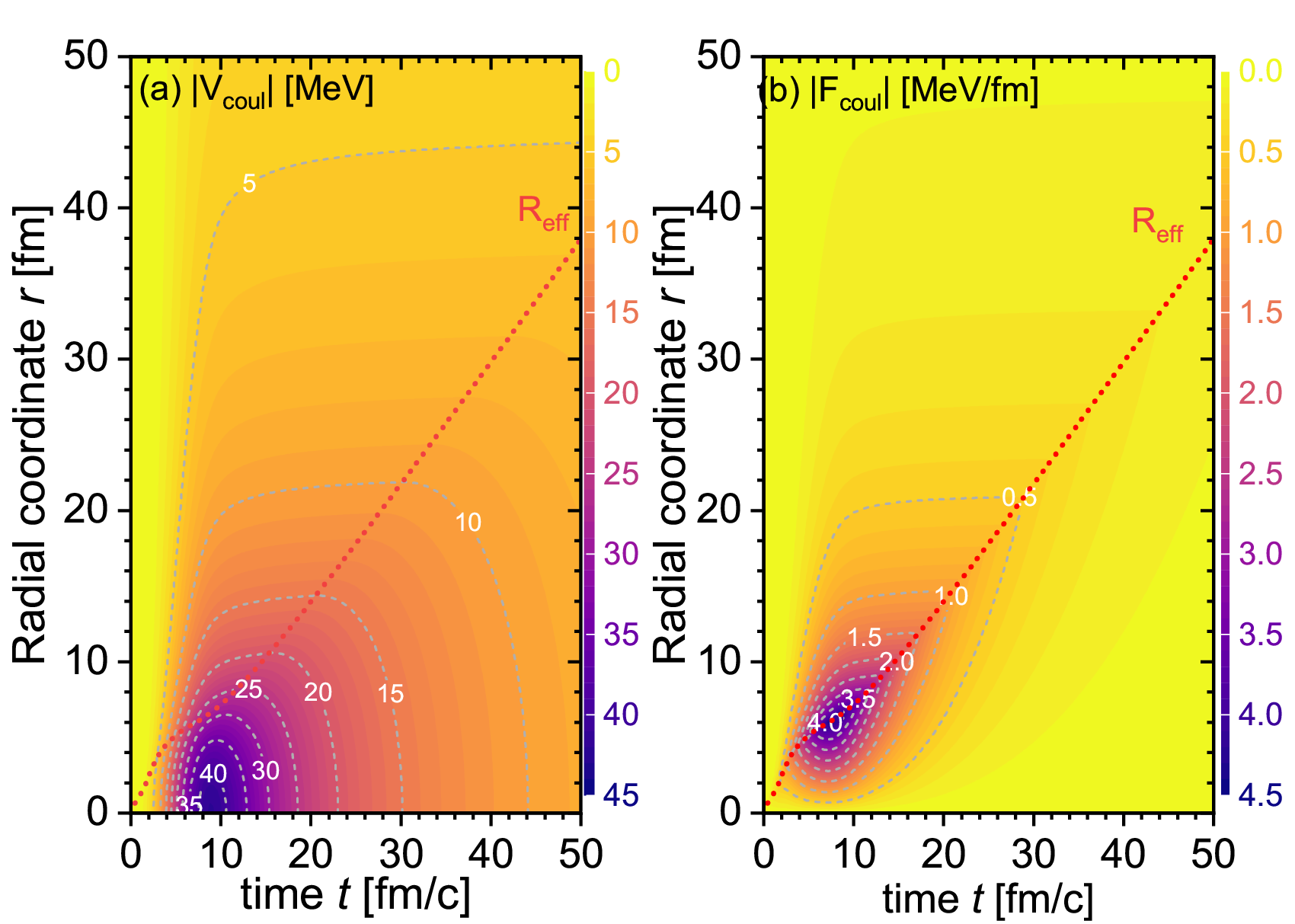}
\caption {\label{fig:time-radius-field}(Color online) Magnitudes of the residual-source Coulomb potential energy (a) and radial force (b) at $\sqrt{s_{NN}}=3.0$ GeV as functions of time and radial coordinate. The red dotted curve indicates $R_{\mathrm{eff}}(t)$.}
\end{figure}


For a unit positive test charge, the Coulomb potential energy generated by the event-averaged uniform sphere is
\begin{equation}
 V(r,t)=\alpha\hbar c Z_{\mathrm{eff}}(t)
 \begin{cases}
 \dfrac{3-r^2/R_{\mathrm{eff}}^2(t)}{2R_{\mathrm{eff}}(t)}, & r<R_{\mathrm{eff}}(t),\\[0.8em]
 \dfrac{1}{r}, & r\geq R_{\mathrm{eff}}(t),
 \end{cases}
\label{eq:potential}
\end{equation}
where $r=|\boldsymbol{r}|$ and $\alpha$ is the fine-structure constant \cite{Gyulassy:1980xb,Baym:1996wk}. 
The potential energy of a pion is $U_{\pi}(r,t)=s_{\pi}V(r,t)$, with $s_{\pi}=+1$ for $\pi^+$ and $s_{\pi}=-1$ for $\pi^-$. 
The corresponding force is
\begin{equation}
 \boldsymbol{F}_{\pi}(\boldsymbol{r},t)=s_{\pi}\alpha\hbar c\,Z_{\mathrm{eff}}(t)
 \begin{cases}
 \dfrac{\boldsymbol{r}}{R_{\mathrm{eff}}^3(t)}, & r<R_{\mathrm{eff}}(t),\\[0.8em]
 \dfrac{\boldsymbol{r}}{r^3}, & r\geq R_{\mathrm{eff}}(t).
 \end{cases}
\label{eq:force}
\end{equation}
Thus, the field repels $\pi^+$ pions radially outward and attracts $\pi^-$ pions radially inward. 
At $r=0$, the force vanishes; inside the sphere, its magnitude increases linearly with $r$; and outside the sphere, it decreases as $1/r^2$. 

Because individual pion freeze-out times are continuous, the reconstructed source parameters are linearly interpolated between adjacent snapshots separated by 0.4 fm/$c$. 
Interpolation allows pions with arbitrary freeze-out times to sample the source evolution continuously. 
As shown in Fig. \ref{fig:zeff-reff-time}, by 50 fm/$c$, $Z_{\mathrm{eff}}(t)$ has nearly saturated, while radial expansion has substantially reduced the field strength. 
Thus, at later times, $Z_{\mathrm{eff}}$ is fixed at its value at 50 fm/$c$, and $R_{\mathrm{eff}}$ is extrapolated linearly.
Pion propagation is continued in steps of 0.4 fm/$c$ to 150 fm/$c$, and the resulting momenta are used as the final momenta in the analysis. 

Figure~\ref{fig:time-radius-field} shows the magnitudes of the residual-source Coulomb potential energy and radial force at $\sqrt{s_{NN}}=3.0$ GeV for central Au+Au. 
The potential energy reaches 42.2 MeV at the origin near $t=8.8$ fm/$c$, during the early compact stage of the source. 
The force reaches a maximum of approximately 4.31 MeV/fm near $t=7.2$ fm/$c$ and $r=6.0$ fm, close to the instantaneous surface of the sphere. 
This spatial separation of the extrema follows directly from the uniform-sphere field: the potential energy is maximal at the center, whereas the force vanishes there, increases linearly inside the source, and decreases as $1/r^2$ outside. 
The rapid weakening of the field at late times reflects the continued increase of $R_{\mathrm{eff}}(t)$ after $Z_{\mathrm{eff}}(t)$ has nearly saturated.

\subsection{Pion-by-pion propagation and correlation analysis}
\label{subsec:propagation}

For each charged pion, the transport calculation provides the last strong-interaction four-coordinate $x_f^{\mu}=(t_f,\boldsymbol{r}_f)$ and four-momentum $p_f^{\mu}=(E_f,\boldsymbol{p}_f)$. 
The subsequent trajectory satisfies
\begin{equation}
 \frac{d\boldsymbol{r}}{dt}=\frac{\boldsymbol{p}}{\sqrt{m_\pi^2+|\boldsymbol{p}|^2}},
 \qquad
 \frac{d\boldsymbol{p}}{dt}=\boldsymbol{F}_{\pi}(\boldsymbol{r},t).
 \label{eq:eom}
\end{equation}
Starting from $\boldsymbol{r}(t_f)=\boldsymbol{r}_f$ and $\boldsymbol{p}(t_f)=\boldsymbol{p}_f$, these equations are integrated using a relativistic velocity-Verlet algorithm \cite{Swope:1982qc,Horowitz:2008vf}. 
A single integration step from $t_n$ to $t_{n+1}=t_n+h$ is given by
\begin{align}
 \boldsymbol{p}_{n+1/2}&=\boldsymbol{p}_n+\frac{h}{2}\boldsymbol{F}_{\pi}(\boldsymbol{r}_n,t_n),\\
 \boldsymbol{r}_{n+1}&=\boldsymbol{r}_n+h\frac{\boldsymbol{p}_{n+1/2}}
 {\sqrt{m_{\pi}^{2}+|\boldsymbol{p}_{n+1/2}|^2}},\\
 \boldsymbol{p}_{n+1}&=\boldsymbol{p}_{n+1/2}+\frac{h}{2}
 \boldsymbol{F}_{\pi}(\boldsymbol{r}_{n+1},t_{n+1}).
 \label{eq:verlet}
\end{align}
Here, $h=0.4$ fm/$c$ and $n$ labels the integration step. 
Each trajectory is propagated from $t_f$ to $t_{\mathrm{stop}}=150$ fm/$c$. A pion with $t_f\geq t_{\mathrm{stop}}$ is left unchanged. 

This procedure modifies both the magnitude and direction of the pion momentum \cite{Pratt:2005hn}. 
Because $F_i\propto s_{\pi}r_i$ for $i=x,y,z$, all three momentum components, $p_x$, $p_y$, and $p_z$, generally change. 
The accumulated Coulomb impulse is
\begin{equation}
 \Delta\boldsymbol{p}_{\pi}=\boldsymbol{p}'_f-\boldsymbol{p}_f
 =\int_{t_f}^{t_{\mathrm{stop}}}\boldsymbol{F}_{\pi}[\boldsymbol{r}(t),t]dt,
 \label{eq:impulse}
\end{equation}
where the prime denotes the momentum after propagation through the residual field.
The polar and azimuthal momentum angles, $\theta_p=\arccos(p_z/|\boldsymbol{p}|)$ and $\phi_p=\operatorname{atan2}(p_y,p_x)$, can both change, as can the angle between $\boldsymbol{p}$ and $\boldsymbol{r}$.
For an exactly radial trajectory, only the signed radial momentum changes.
For a nonradial trajectory, the force deflects the pion within its orbital plane.
The central force conserves $\boldsymbol{r}\times\boldsymbol{p}$ in the continuous equations, whereas the explicit time dependence of $U_\pi$ means that the pion's kinetic plus potential energy need not be constant. 
An initially outward-moving $\pi^+$ is accelerated, whereas an initially outward-moving $\pi^-$ is decelerated. 

After propagation, the phase-space point supplied to CRAB is
\begin{equation}
 x_{\mathrm{CRAB}}^{\mu}=x_f^{\mu},\qquad
 p_{\mathrm{CRAB}}^{\mu}=
 \left(\sqrt{m_{\pi}^2+|\boldsymbol{p}'_f|^2},\boldsymbol{p}'_f\right).
 \label{eq:crab-input}
\end{equation}
Thus, the microscopic freeze-out time and position are retained in the CRAB emission function, whereas the on-shell final momentum includes the accumulated residual Coulomb impulse. 
This construction isolates the momentum-space lensing induced by the residual field. 
Replacing $x_f^{\mu}$ with the propagated asymptotic coordinate would redefine the emission geometry and is therefore not adopted here.

The analysis proceeds as follows. 
First, the collision-energy-dependent source evolution is specified. 
Each pion is then initialized at its individual freeze-out four-coordinate and propagated under the charge-dependent vector force. 
Its on-shell energy is reconstructed, the analysis selections are applied, and like-sign pion pairs are subsequently constructed and analyzed with CRAB \cite{Pratt:1990zq,Pratt:1986ev}. 
All pair observables are evaluated using the modified momenta. 
In particular, $k_T=\frac{1}{2}|\boldsymbol{p}'_{T,1}+\boldsymbol{p}'_{T,2}|$, 
and both the LCMS boost and the Bertsch--Pratt axes are recalculated after propagation: $\boldsymbol{e}_{\mathrm{out}}=\boldsymbol{P}'_T/|\boldsymbol{P}'_T|$, $\boldsymbol{e}_{\mathrm{side}}=\boldsymbol{e}_z\times\boldsymbol{e}_{\mathrm{out}}$, and $\boldsymbol{e}_{\mathrm{long}}=\boldsymbol{e}_z$, where $\boldsymbol{P}'=\boldsymbol{p}'_1+\boldsymbol{p}'_2$. 

Consequently, even though the freeze-out coordinates are retained, their out, side, and long projections onto the momentum-defined axes can change after Coulomb deflection. Four physically distinct Coulomb ingredients enter at different stages of the analysis. 
First, Cases I--III describe the microscopic baryon--baryon, baryon--meson, and meson--meson Coulomb interactions during the UrQMD evolution. 
Second, the Coulomb final-state interaction of each selected pion pair is included in CRAB when the correlation function is constructed~\cite{Koonin:1977fh,Pratt:1990zq,Lisa:2005dd}. 
Third, in the complementary residual-source calculation, each pion from the Case-I phase-space output is propagated from its individual kinetic-freeze-out phase-space point through the time-dependent field specified by $Z_{\rm eff}(t)$ and $R_{\rm eff}(t)$ before pair construction. 
Fourth, the Coulomb correction factor used in the HBT-radius fit is derived from CRAB correlation functions calculated from the same emission sample and pair selection with and without the Coulomb interaction, and it is almost numerically consistent with the HADES finite-source factor over the fit-relevant momentum region \cite{HADES:2019lek}. 
The construction of the three-dimensional correlation function and the Gaussian extraction of $R_{\mathrm{out}}$, $R_{\mathrm{side}}$, and $R_{\mathrm{long}}$ follow Refs.~\cite{Li:2022iil,Li:2022icu,HADES:2019lek,Luong:2024eaq,Luong:2026ugh,STAR:2020dav,HADES:2018gop} and are not repeated here.

\section{Results and discussion}\label{sec:results}
\subsection{Transverse-momentum dependent charged-pion ratio}

\begin{figure}[b!]
\centering
\includegraphics[width=0.45\textwidth]{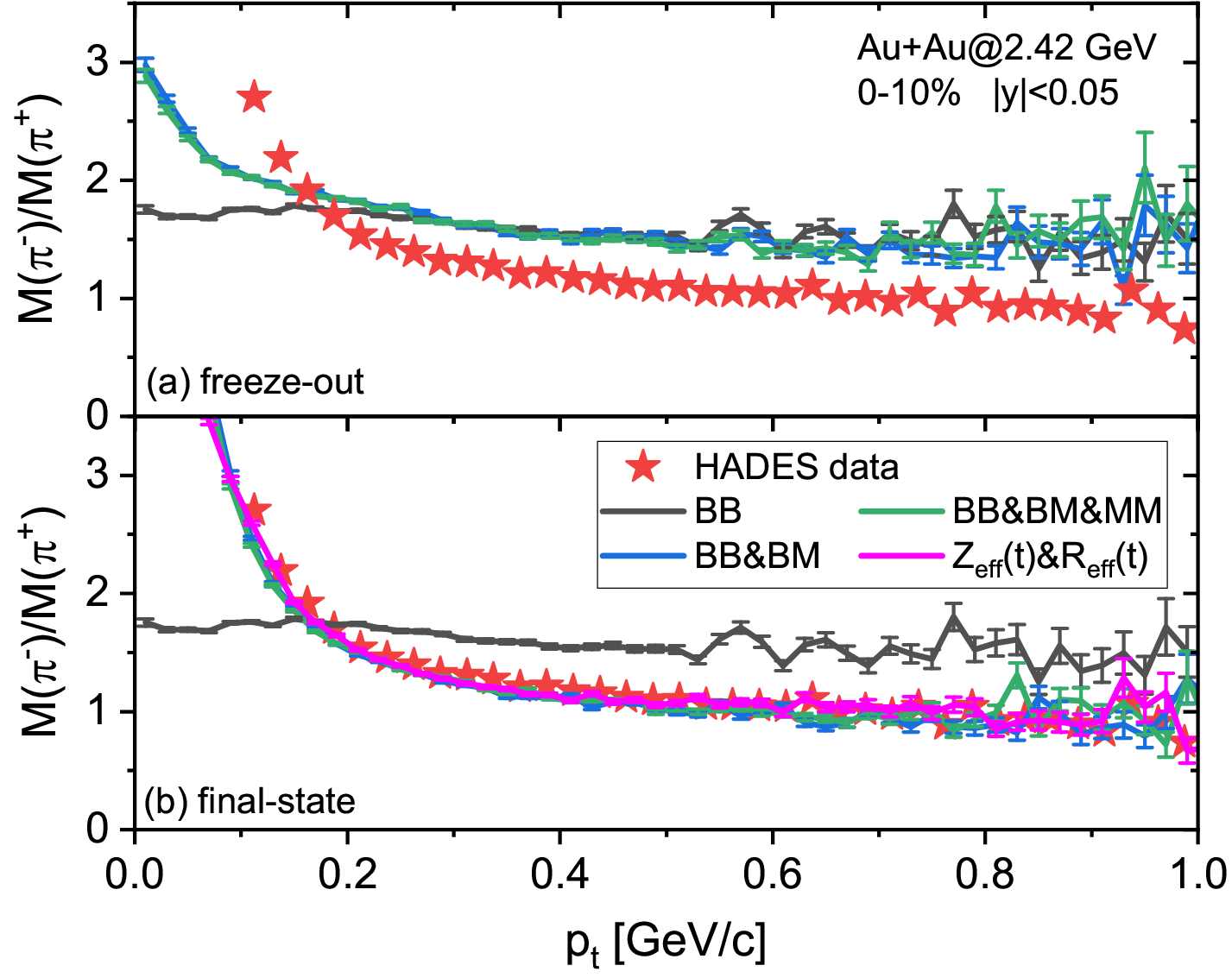}
\caption {\label{afig1_piplus_piminus_ratio_pt_data}(Color online) Transverse-momentum $p_{T}$ dependence of the charged-pion ratio at freeze-out (a) and in the final state (b) for 0--10\% central Au+Au collisions at $\sqrt{s_{NN}}=2.42$ GeV and $|y|<0.05$. Red stars denote the HADES data \cite{HADES:2020ver}. Gray, blue, and green curves represent Cases I (BB), II (BB\&BM), and III (BB\&BM\&MM), respectively. The magenta curve in panel (b) shows the dynamical residual-source calculation based on the Case-I phase-space
output.
}
\end{figure}

Before discussing the femtoscopic observables, the effectiveness of the Coulomb interactions should be examined first. 
Since the transverse momentum $p_{T}$ dependence of the ratio of charged pions is sensitive to the Coulomb potential \cite{Steinheimer:2026xeg,HADES:2022mwn}, Fig.~\ref{afig1_piplus_piminus_ratio_pt_data} depicts the charged pion ratio as a function of $p_{T}$ for central Au+Au collisions with different Coulomb interactions at $\sqrt{s_{NN}}=2.42$ GeV. 
In the top panel, which was calculated using the momentum from the last interaction, the BB calculation exhibits a much weaker $p_T$ dependence and does not reproduce the pronounced low-$p_T$ enhancement. 
Once the direct BM Coulomb interaction is included, the ratio develops a strong low-$p_T$ rise already at freeze-out.
The close agreement between the BB\&BM and BB\&BM\&MM calculations indicates that the additional MM interaction has a comparatively small influence on this spectral ratio. 

After the subsequent transport evolution, the BM-induced charge separation is further manifested in the final-state spectra [Fig.~\ref{afig1_piplus_piminus_ratio_pt_data}(b)]. 
The dynamical residual-source calculation generates a similar low-$p_T$ enhancement even though it starts from the Case-I phase-space output. 
Since negatively charged pions are attracted and shifted toward lower momenta, whereas positively charged pions are repelled and shifted toward higher momenta. 
The corresponding momentum modification is strongest for early-freezing pions and decreases progressively with freeze-out time as the charged source expands and its Coulomb field weakens. 
Consequently, the $\pi^-/\pi^+$ ratio is enhanced at low $p_T$ and reduced at larger $p_T$. 
The microscopic BM calculation and the effective residual-source calculation describe the main trend of the measured ratio over the whole $p_T$ region. 
However, the similar single-particle ratios obtained with the two Coulomb treatments do not imply equivalent emission functions: their different modifications of the pion trajectories and freeze-out position–momentum correlations are tested more directly by the HBT observables below.

\subsection{Pair transverse momentum dependent HBT radii}

\begin{figure}[t!]
\centering
\includegraphics[width=0.45\textwidth]{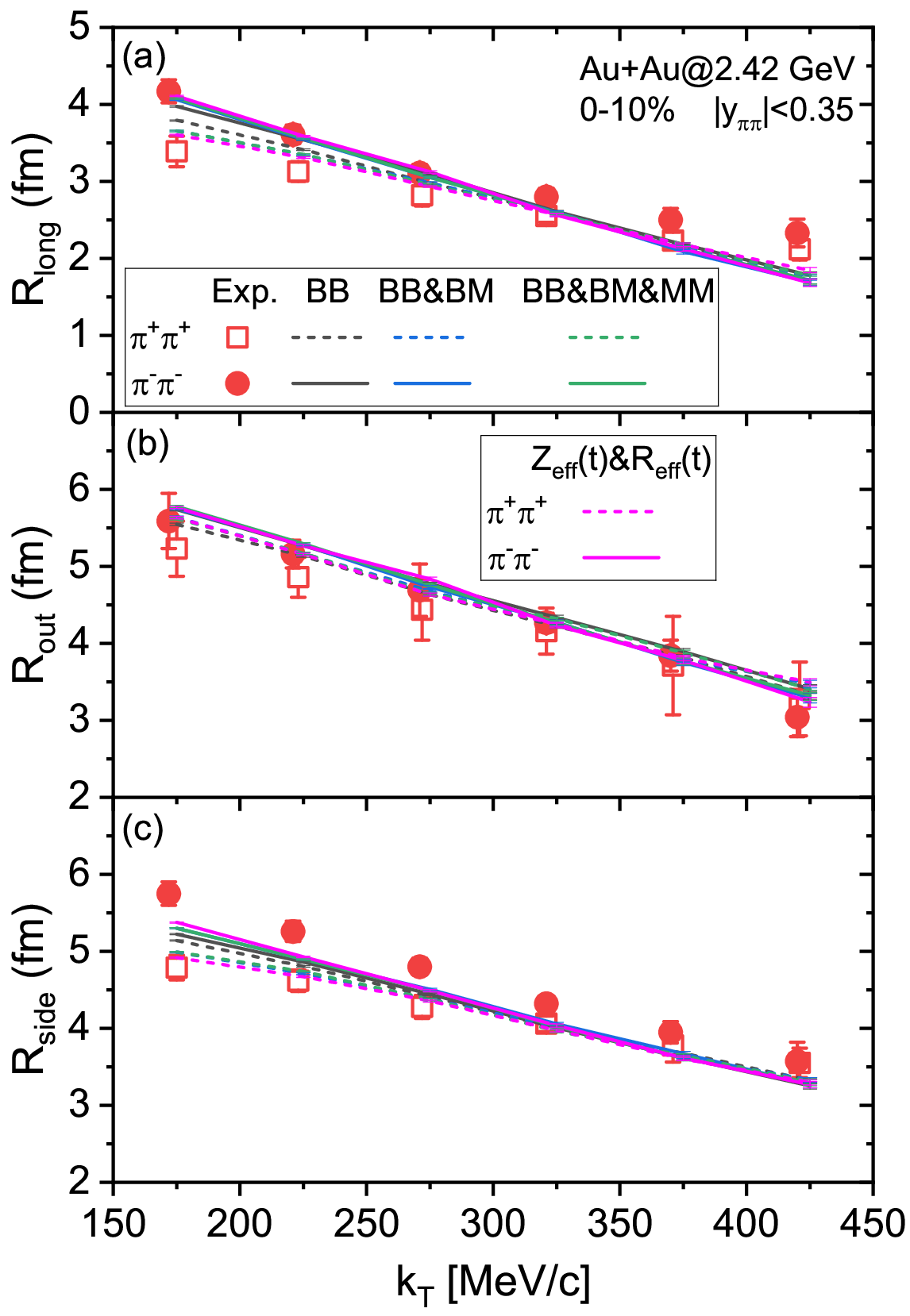}
\caption {\label{fig3_HBT_R_Kt_242GeV}(Color online)
HBT radii $R_{\mathrm{long}}$ (a), $R_{\mathrm{out}}$ (b), and $R_{\mathrm{side}}$ (c) as functions of $k_T$ in 0--10\% central Au+Au collisions at $\sqrt{s_{NN}}=2.42$ GeV. Open squares and filled circles represent the measured $\pi^+\pi^+$ and $\pi^-\pi^-$ radii, respectively \cite{HADES:2019lek}. Dotted and solid curves show the corresponding calculations. The line colors have the same meanings as in Fig.~\ref{afig1_piplus_piminus_ratio_pt_data}(b).
}
\end{figure}

Then, the calculated and measured HBT radii at $\sqrt{s_{NN}}=2.42$ GeV are compared and shown in Figure~\ref{fig3_HBT_R_Kt_242GeV}. 
All calculations reproduce the pronounced decrease of $R_{\mathrm{long}}$, $R_{\mathrm{out}}$, and $R_{\mathrm{side}}$ with increasing $k_T$. 
This common trend reflects the smaller homogeneity regions probed by higher-$k_T$ pion pairs and is considerably stronger than the differences among the Coulomb scenarios \cite{Wiedemann:1999qn,Heinz:1999rw,Lisa:2005dd,STAR:2014shf}. 
The calculated radii are also of the same overall magnitude as the experimental values. 

The comparison of Cases I--III isolates the effects of microscopic two-body Coulomb interactions. 
At $k_T=175$ MeV/$c$, including BM Coulomb interactions increases $R_{\mathrm{long}}^{\pi^{-}\pi^{-}}/R_{\mathrm{long}}^{\pi^{+}\pi^{+}}$ from 1.048 in Case I to 1.110 in Case II, 
and increases $R_{\mathrm{side}}^{\pi^{-}\pi^{-}}/R_{\mathrm{side}}^{\pi^{+}\pi^{+}}$ from 1.016 to 1.062. 
The additional MM interaction changes these ratios only to 1.115 and 1.062, respectively. 
The same hierarchy persists near $k_T=225$ MeV/$c$. 
Direct Coulomb coupling between pions and charged baryons therefore accounts for most of the additional microscopic splitting at low $k_T$, whereas the MM contribution is comparatively small.

The residual-source calculation produces the largest longitudinal and sideward splitting among the calculated scenarios at low $k_T$.
At $k_T=175$ MeV/$c$, it yields ratios of 1.128, 1.024, and 1.085 for the long, out, and side components, respectively. 
The corresponding experimental ratios are approximately 1.23, 1.07, and 1.20. 
The residual-source field shifts the long and side radii toward the observed ordering without reproducing the full magnitude of the splitting, and this effect decreases rapidly with increasing $k_T$.
The $R_{\mathrm{out}}$ ratio remains much closer to unity over the full $k_T$ range, consistent with the weaker experimental charge ordering and larger uncertainties for this component.

\subsection{Collision-energy dependent HBT radii}

\begin{figure}[t!]
\centering
\includegraphics[width=0.45\textwidth]{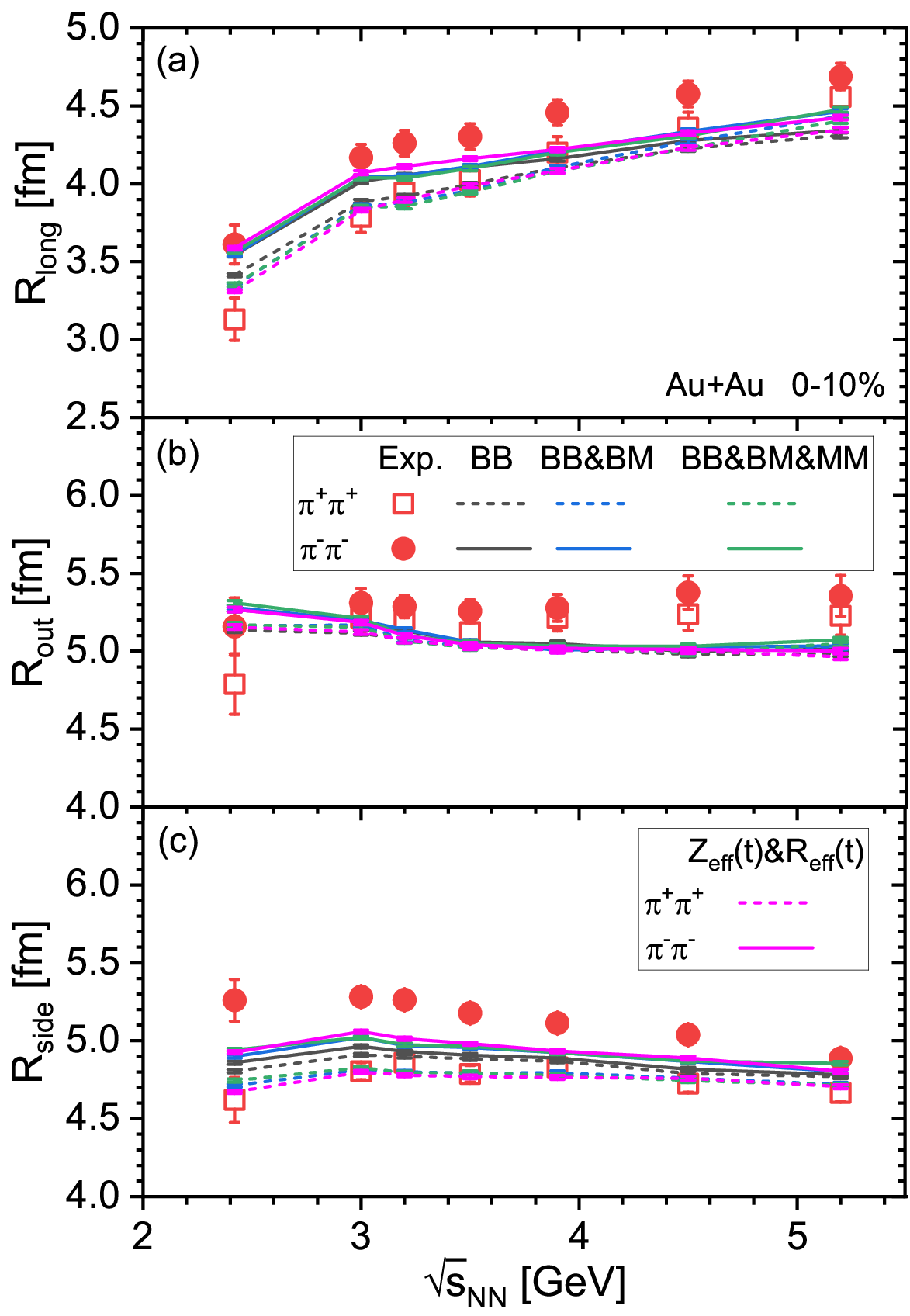}
\caption {\label{fig4_radii_energy}(Color online)
Collision-energy dependence of $R_{\mathrm{long}}$ (a), $R_{\mathrm{out}}$ (b), and $R_{\mathrm{side}}$ (c) in 0--10\% central Au+Au collisions at $k_T=200\pm25$ MeV/$c$ and $|y_{\pi\pi}|<0.5$, while $k_T=225\pm25$ MeV/$c$ and $|y_{\pi\pi}|<0.35$ are used for $\sqrt{s_{NN}}$=2.42 GeV. 
Experimental data are from HADES and STAR \cite{HADES:2019lek,Luong:2024eaq,Luong:2026ugh}. The line styles and colors have the same meanings as in Fig.~\ref{fig3_HBT_R_Kt_242GeV}.
}
\end{figure}

Next, the excitation functions at $k_T=200\pm25$ MeV/$c$ ($k_T=225\pm25$ MeV/$c$ for $\sqrt{s_{NN}}$=2.42 GeV) are shown in Fig.~\ref{fig4_radii_energy}. 
The calculated $R_{\mathrm{long}}$ increases from approximately 3.3--3.6 fm at 2.42 GeV to 4.3--4.5 fm at 5.2 GeV, following the measured increase for both charge states. 
In contrast, $R_{\mathrm{out}}$ remains close to 5 fm, and $R_{\mathrm{side}}$ changes only moderately over the same collision-energy interval. 
Thus, among the three components, the longitudinal homogeneity length exhibits the strongest collision-energy dependence, whereas the transverse homogeneity lengths vary more slowly.

At 2.42 GeV, the residual-source calculation gives $R_{\mathrm{long}}^{\pi^{+}\pi^{+}}=3.31$ fm and $R_{\mathrm{long}}^{\pi^{-}\pi^{-}}=3.59$ fm, compared with the measured values of 3.13 and 3.61 fm, respectively. 
The corresponding calculated side radii are 4.67 and 4.93 fm, compared with the measured values of 4.62 and 5.26 fm. 
At this energy, the positive-pion side radius lies close to the measured value, whereas the negative-pion side radius remains underestimated. 
The remaining discrepancy in the sideward charge splitting is thus associated primarily with the negative-pion homogeneity length rather than with a common offset of both charge states.

The differences among Cases I--III are substantially smaller than the overall collision-energy dependence of $R_{\mathrm{long}}$. 
Nevertheless, the BM Coulomb interaction systematically increases the negative-pion long and side radii relative to the corresponding positive-pion radii at low collision energies. 
The residual-source field produces a similar but somewhat larger shift. 
Toward 5.2 GeV, both the calculated charge-separated curves and the experimental data points converge. 
The model underestimates several absolute radii at higher collision energies, particularly $R_{\mathrm{out}}$, demonstrating that agreement in a charge ratio does not necessarily imply a complete description of the underlying space--time emission structure.

\subsection{Collision-energy dependent HBT-radius ratios}

\begin{figure}[t!]
\centering
\includegraphics[width=0.45\textwidth]{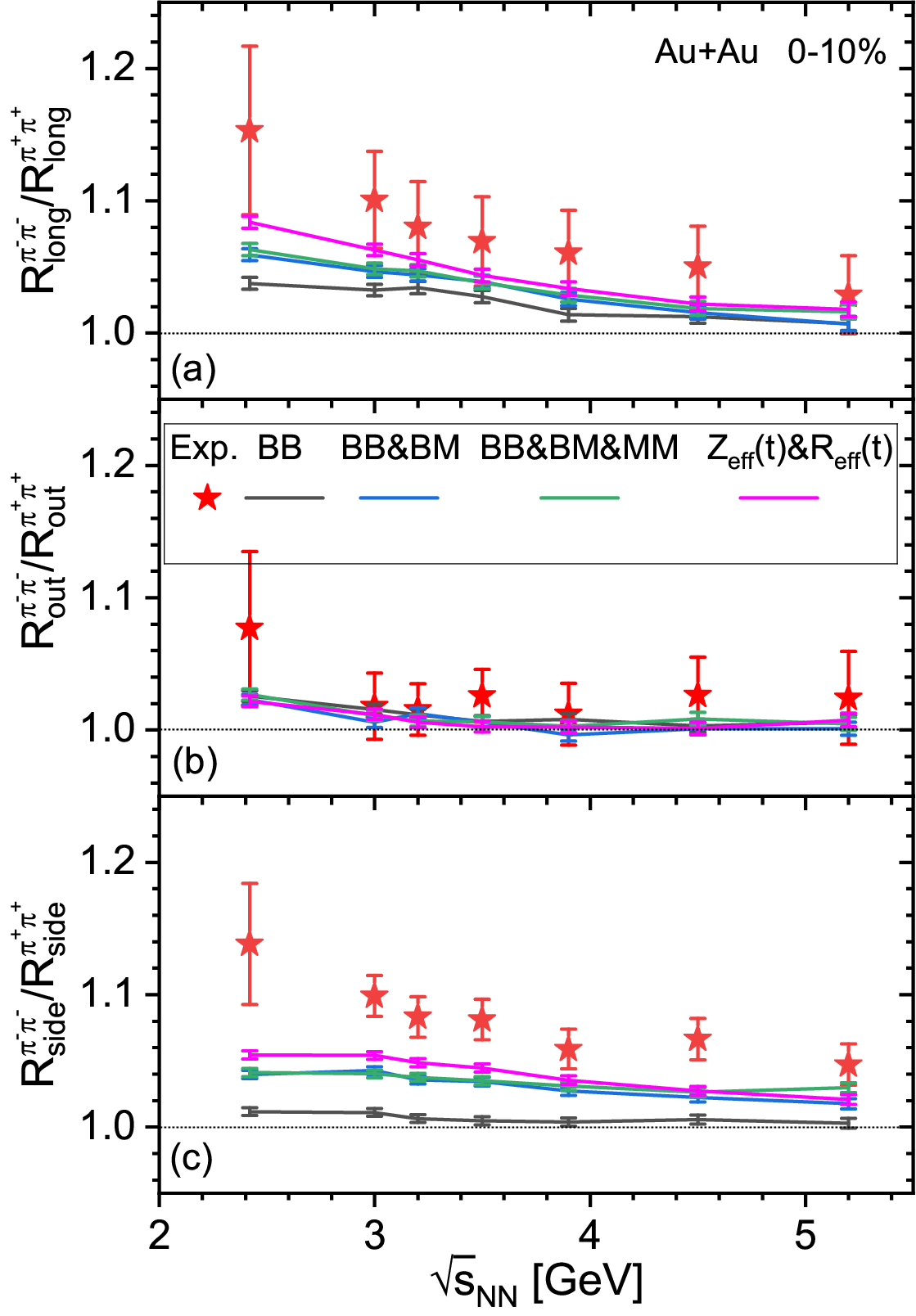}
\caption {\label{fig5_radii_Ratio_Snn}(Color online)
Collision-energy dependence of $\mathcal R_i=R_i^{\pi^{-}\pi^{-}}/R_i^{\pi^{+}\pi^{+}}$ for the long (a), out (b), and side (c) components in 0--10\% central Au+Au collisions. Stars represent the experimental ratios from HADES and STAR \cite{HADES:2019lek,Luong:2024eaq,Luong:2026ugh}. The black, blue, and green curves show Cases I--III, respectively. The magenta curve denotes the dynamical residual-source calculation based on Case I. The horizontal dotted line indicates unity.
}
\end{figure}

At last, to clearly show the effect of the dynamical Coulomb interactions on the charge splitting in HBT radii between charged pions, Figure~\ref{fig5_radii_Ratio_Snn} summarizes the charge splitting in terms of $\mathcal R_i=R_i^{\pi^{-}\pi^{-}}/R_i^{\pi^{+}\pi^{+}}$. 
The experimental $\mathcal R_{\mathrm{long}}$ and $\mathcal R_{\mathrm{side}}$ ratios are largest at 2.42 GeV and approach unity with increasing collision energy. 
The calculated ratios exhibit the same overall trend. 
In both the data and the calculations, $\mathcal R_{\mathrm{out}}$ remains close to unity and shows no comparably strong or persistent charge ordering.

Case I already yields ratios slightly above unity. 
The reference retains charge-dependent hadronic channels, the neutron-rich composition of Au, and the indirect effects of BB Coulomb propagation; their individual contributions are not isolated here. 
These ingredients are kept fixed rather than varied separately. 
At 2.42 GeV, including BM Coulomb propagation increases $\mathcal R_{\mathrm{long}}$ from 1.038 to 1.059 and $\mathcal R_{\mathrm{side}}$ from 1.012 to 1.040. 
The additional MM propagation changes these ratios only to 1.063 and 1.041, respectively. 
This comparison indicates that direct pion--baryon Coulomb coupling is the dominant microscopic Coulomb contribution at low collision energies.

Applied independently to Case I, the dynamical residual-source field increases the 2.42 GeV ratios to $\mathcal R_{\mathrm{long}}=1.084$ and $\mathcal R_{\mathrm{side}}=1.054$. 
At 3.0 GeV, the residual-source calculation gives 1.063 and 1.054, compared with the experimental values of 1.101 and 1.099. At 5.2 GeV, the calculated ratios decrease to 1.018 and 1.021, whereas the measured ratios are 1.029 and 1.047. 
This collision-energy dependence is consistent with a stronger Coulomb imprint when the residual source remains compact for a longer interval. 
Nevertheless, the sideward splitting remains underestimated over much of the investigated energy range.

The Case III and residual-source curves are not expected to coincide, although both account for pion coupling to the charged environment. 
The microscopic calculation resolves discrete charges and event-by-event anisotropies throughout the transport evolution, whereas the residual-source calculation retains only a smooth monopole field after pion freeze-out. 
Moreover, the mapping from the source parameters to the trajectory-integrated Coulomb impulse is nonlinear. 
Consequently, the force calculated from the averaged $Z_{\mathrm{eff}}(t)$ and $R_{\mathrm{eff}}(t)$ is generally not equal to the event average of the microscopic forces sampled along correlated pion trajectories. 
Case III can also modify the freeze-out coordinates and the position--momentum correlations that determine the HBT radii, whereas the residual-source calculation retains the Case-I emission coordinates and passes the propagated momenta to CRAB. 
The difference between the two calculations, therefore, reflects the sensitivity to the dynamical representation and temporal extent of the Coulomb field \cite{Hardtke:1997cy,Pratt:2005hn,Wei:2021yiy}. 

Taken together, the two analyses show that both microscopic and residual-source Coulomb dynamics can generate finite $\pi^-\pi^-$--$\pi^+\pi^+$ splitting in the extracted HBT radii. 
They do not, however, establish Coulomb interactions as the sole origin of the measured splitting.
Possible sources of the remaining discrepancy include isospin-dependent pion production and absorption, charge-dependent hadronic potential, resonance decay, nonspherical source geometry, and event-by-event charge-density fluctuations \cite{Jiang:2026pox,Kincses:2025iaf,Steinheimer:2026xeg,Khyzhniak:2026skh}. 
Their interplay with Coulomb propagation, together with a more detailed charged-source geometry, warrants further study. 
Within the present framework, Coulomb dynamics make a non-negligible contribution to the charge-dependent HBT radii at low collision energies.

\section{Summary and outlook}\label{sec:summary}

We have developed a UrQMD+CRAB framework that treats microscopic two-body Coulomb interactions separately from the post-freeze-out Coulomb field of the residual charged source. 
Three transport scenarios distinguish the effects of baryon--baryon, baryon--meson, and meson--meson Coulomb interactions. 
As a dynamical extension of effective-source descriptions, the present approach reconstructs the expanding residual source in terms of $Z_{\mathrm{eff}}(t)$ and $R_{\mathrm{eff}}(t)$. 
The charge-dependent force is then integrated along the trajectory of each pion from its individual last strong-interaction point.

The calculations reproduce the dominant decrease of all three HBT radii with increasing $k_T$ and the increase of $R_{\mathrm{long}}$ with collision energy. 
Baryon--meson Coulomb interactions generate most of the additional microscopic charge splitting, whereas the meson--meson contribution is smaller. 
When applied to the Case-I reference, the dynamical residual-source field enhances $R_{\mathrm{long}}^{\pi^{-}\pi^{-}}/R_{\mathrm{long}}^{\pi^{+}\pi^{+}}$ and $R_{\mathrm{side}}^{\pi^{-}\pi^{-}}/R_{\mathrm{side}}^{\pi^{+}\pi^{+}}$, and both ratios approach unity as $\sqrt{s_{NN}}$ increases. 
The $R_{\mathrm{out}}$ ratio is only weakly affected.
These results demonstrate that the evolving Coulomb environment can account for part of the charged-pion HBT-radius splitting, but the observed charged splitting is not reproduced quantitatively in all three directions, and the $R_{\mathrm{side}}$ ratio is most consistently underestimated. 


Future extensions may incorporate event-by-event, nonspherical charge distributions and explicitly correlate the freeze-out space--time point of each surviving pion with its subsequent propagation in the evolving electromagnetic field. 
Such a trajectory-resolved analysis would determine when and where the Coulomb momentum shift is accumulated and clarify how the resulting charge splitting is coupled to nuclear structure, the nuclear equation of state, isospin-dependent dynamics, and resonance dynamics \cite{Jiang:2026pox,Kincses:2025iaf,Steinheimer:2026xeg,Khyzhniak:2026skh}. 
This will help disentangle electromagnetic acceleration from strong-interaction modifications of the emission geometry and establish the sensitivity of charge-dependent HBT observables to the space--time evolution of net charge in baryon-rich matter.

\section*{Acknowledgements}
We thank Dr. Jan Steinheimer and Prof. Marcus Bleicher for helpful discussions and valuable comments on the manuscript.
This work was supported in part by 
the National Natural Science Foundation of China under Grant Nos. 12505143, 12335008, and 12675166, 
the National Key Research and Development Program of China under Grant No. 2023YFA1606402, and 
the Zhejiang Provincial Natural Science Foundation of China under Grant No. LQN25A050003.
The authors are grateful to the C3S2 computing center at Huzhou Normal University for computational support.
P.~C.~Li gratefully acknowledges financial support from the China Scholarship Council under Grant No.~202608330358.

\end{document}